\documentclass[12pt]{article}
\usepackage[utf8]{inputenc}
\usepackage{graphicx}
\usepackage{subcaption}
\usepackage{array}
\usepackage{amssymb}
\usepackage{amsfonts}
\usepackage{amsmath}
\usepackage{mathrsfs}
\usepackage{color}
\usepackage{booktabs}
\usepackage{threeparttable}
\usepackage{multirow}
\usepackage{times}
\usepackage{epsfig}
\usepackage{chngpage}
\usepackage{latexsym}
\usepackage{mathrsfs}
\usepackage{bm}
\usepackage{epstopdf}

\usepackage{xurl}
\usepackage{cite}
\usepackage[english]{babel}

\usepackage[a4paper,top=2cm,bottom=2cm,left=3cm,right=3cm,marginparwidth=1.75cm]{geometry}

\usepackage{changes}
\usepackage{authblk}
\usepackage{amsmath}
\usepackage{graphicx}
\usepackage[colorlinks=true, allcolors=blue]{hyperref}
\newif\ifshowdeleted
  \showdeletedfalse   % Hide deleted content

\let\olddeleted\deleted
\renewcommand{\deleted}[1]{%
  \ifshowdeleted\olddeleted{#1}\else\ignorespaces\fi}

\begin{document}
% \fontsize{15}{18}\selectfont

\title{The landscape of problematic papers in the field of non-coding RNA}
 % \author{anonymous}
\author[1,2]{Ying Lou}
\author[1,2]{Zhengyi Zhou}
\author[3]{Guosheng Wang}
\author[1,2]{Zhesi Shen\thanks{Corresponding author: shenzhs@mail.las.ac.cn (ZS)}}
\author[1,2]{Menghui Li\thanks{Corresponding author: limh@mail.las.ac.cn (ML)}}
\affil[1]{National Science Library, Chinese Academy of Sciences, Beijing 100190, P. R. China}
\affil[2]{Department of Information Resources Management, School of Economics and Management, University of Chinese Academy of Sciences, Beijing 100190, P. R. China}
\affil[3]{Suzhou Collaborative Open Science Research Center (SCOS), Suzhou 215008, Jiangsu, P. R. China}
\date{}
\maketitle

\begin{abstract}
Retractions have increased sharply in recent years, alongside a growing number of papers that receive post-publication comments questioning their reliability (commented papers). Together, retracted and commented papers undermine the credibility of scientific research and may also threaten public health. In this study, we examine problematic papers in the field of non-coding RNA (ncRNA) from multiple perspectives to identify common patterns and inform strategies for addressing large-scale fraudulent publications. We find that studies on under-investigated ncRNAs are more likely to become problematic papers. These papers often show substantial textual similarity, and many additional papers with similar text also display suspicious image duplication. Healthcare institutions, particularly those with lower publication output, appear especially vulnerable to producing such papers. Most problematic papers are concentrated in a small set of journals, many of which do not adequately address concerns raised after publication. Overall, our findings indicate that a substantial number of problematic papers may remain undetected and that their shared characteristics can support more effective strategies for identifying and curbing large-scale fraudulent publications.

\textbf{Keywords:} Research Integrity; Fraudulent Paper; Textual Similarity; Non-coding RNA; Paper Mill; 
\end{abstract}

\section{Introduction}

Retractions have increased markedly in recent years, drawing broad attention to the reliability of the scientific record \cite{Nature:NEWRECORD,JDIS:Amend}. At the same time, evidence suggests that fraudulent publication, including work produced by organized paper mills, is expanding faster than the growth of legitimate research output \cite{PNAS:editor}. Research integrity has therefore become a central concern across the scholarly community \cite{PNAS:Misconduct}.

Formal retractions, however, capture only part of the problem. A much larger body of literature has been questioned on post-publication peer-review platforms such as PubPeer, where comments often focus on image problems, implausible data, or other indicators of unreliability \cite{PNAS:editor, JASIST:PubPeer}. Although retracted papers are explicitly marked as unreliable, papers that receive serious post-publication criticism (hereafter, commented papers) may pose a comparable risk, especially because they are far more numerous \cite{JASIST:PubPeer,bioRxiv:ncRNA}. Taken together, retracted and commented papers constitute a broader category of problematic papers that may continue to shape subsequent research even after concerns have been raised \cite{RIPR:misbehaviors,RIPR:Spread}.

The consequences of this problematic literature are substantial. When unreliable findings remain in circulation, later studies may build on flawed assumptions, clinicians and policymakers may be misled, and public trust in science may be weakened \cite{JAMA:citation,QSS:policy,PNAS:cross-platform,BJU:urology}. These effects are especially concerning in the life and clinical sciences, which account for a large share of retractions \cite{INNOVATION:Sciencemap}. Retracted studies in these fields have even been incorporated into systematic reviews and meta-analyses, thereby compromising evidence-based medicine \cite{BMJ:evidence,JAMA:SR,JAMAMedicine:SR}. Understanding how problematic papers emerge and cluster in such fields is, therefore, an important research integrity task.

Within this broader context, non-coding RNA (ncRNA) is a particularly important case. The field has expanded rapidly, generating more than $153,000$ publications since 2000 \cite{NRG:ncRNA}. At the same time, it shows unusually high rates of both retraction ($1.92\%$) and post-publication comment ($5.92\%$), far above the cross-disciplinary averages of $0.08\%$ and $0.27\%$, respectively \cite{bioRxiv:ncRNA}. Problematic ncRNA papers have also been cited in patents and clinical trials, extending their potential influence beyond academic publishing \cite{bioRxiv:ncRNA}. This vulnerability is compounded by a major knowledge imbalance: only about $2,000$ of more than $100,000$ human ncRNA genes have been studied extensively \cite{PLOSBIOLOGY:under-investigate,NRMC:lncRNA}. In a rapidly growing yet unevenly studied field, limited domain familiarity may reduce scrutiny and create favorable conditions for paper mills and other forms of low-quality or fraudulent research \cite{NAUNYN:nucleotide,BIOMARKER:fraudulent}.

Despite the severity of this problem, systematic analyses remain limited. Existing studies on retracted papers have generated important insights into retraction trends, integrity indicators, post-retraction citations, gender disparities, and sanctions imposed on researchers \cite{Nature:NEWRECORD,JDIS:Amend,arxiv:RI2,INNOVATION:Sciencemap,PLOSONE:Self-correction,RIPR:Spread,JOI:sentiment,JOI:gender,JOI:penalties}. They have also examined the downstream effects of retracted work on scientific development, technology, altmetrics, and funding \cite{ACCOUNTABILITY:hematology,SCIENTOMETRICS:framework}. Current research frameworks emphasize the harms associated with retracted papers \cite{eLife:Financialcosts,PNAS:cross-platform,QSS:policy}. Nevertheless, two gaps remain. First, problematic papers have rarely been studied as a broader set that includes both retracted and commented papers \cite{PNAS:editor}. Second, scalable methods for detecting suspicious clusters of papers are still underdeveloped, even though prior work has shown that questionable publications often display strong textual similarity \cite{SCIENTOMETRICS:similarities}. Recent mapping approaches based on PubMedBERT and t-SNE provide a promising way to examine such similarity patterns across the biomedical literature \cite{Patterns:landscape,SCIENTOMETRICS:similarities}.

Addressing these gaps also requires attention to the broader publishing ecosystem. Problematic papers are not produced in isolation: they are shaped by research incentives, institutional evaluation systems, editorial practices, and publishing models. Some institutions employ inappropriate metrics that reward quantity over quality, leaving researchers vulnerable to exploitation by paper mills \cite{PNAS:editor,NATURE:papermill,NATURE:CASHREWARDS}. Meanwhile, the rise of Open Access journals prioritizing volume over quality control has created a permissive environment \cite{PEERJ:mega-journal}. In this context, paper mills play a critical role in exploiting these gaps, channeling fraudulent manuscripts into suitable outlets \cite{PNAS:editor}. A field-level analysis should therefore examine not only the content of problematic papers but also the topics they target, the institutional settings from which they emerge, and the journals in which they concentrate.

Against this background, this study provides a comprehensive analysis of problematic papers in the ncRNA field. We ask four questions: Which ncRNA topics are disproportionately associated with problematic papers? To what extent do these papers exhibit textual similarity? What kinds of institutions are most frequently associated with them? Which journals publish the largest concentrations of such work? By answering these questions, we aim to clarify the structure of problematic publications in ncRNA and to provide an empirical basis for more effective detection and prevention strategies.

\section{Materials and Methods}
We designed the analysis to address four questions introduced above: the topical concentration of problematic papers, their textual similarity, the institutional settings from
which they emerged, and the journals in which they were concentrated. To do so, we first constructed the ncRNA publication corpus, then linked it to retraction and post-
publication-comment data, identified named microRNAs and lncRNAs in titles and abstracts, projected papers into a published biomedical similarity map, assessed potential
image duplication, and classified institutions by organization type.

\subsection{Data}
\textit{ncRNA corpus}. The ncRNA corpus was retrieved from Clarivate’s InCites using the citation topics classification system, which clusters Web of Science records using a Leiden algorithm based on citation linkages. This system assigns each article to a three-level topic hierarchy consisting of macro-topics, meso-topics, and micro-topics. Macro-topics and meso-topics are manually annotated, whereas micro-topics are labeled automatically using the most salient keywords in each cluster. Additional information on this topic-classification system is available from Clarivate’s documentation: \url{https://incites.zendesk.com/hc/en-gb/articles/22514077746961-Citation-Topics}.

On January 3, 2025, we retrieved approximately $153,900$ records assigned to the meso-topic ``Micro \& Long Noncoding RNA". The dataset covered publications from 2000 to 2023 and included articles, reviews, and retracted publications. Within this meso-topic, papers were further distributed across four micro-topics: MicroRNA in Cancer, lncRNA, Exosomes, and RNA interference (RNAi) (Table \ref{tab:microtopics}). We also extracted metadata on author affiliations, journals, and publishers from Web of Science.

\textit{Retracted and commented papers}.To identify problematic papers, we combined two external sources. First, we retrieved $2,961$ retracted papers from the Amend platform \cite{JDIS:Amend}. Second, we identified $9,108$ commented papers from PubPeer by matching ncRNA records through DOI or PMID. The integrated dataset is available at: \url{https://zenodo.org/doi/10.5281/zenodo.13383979} \cite{bioRxiv:ncRNA}.  Among these records, $2,617$ papers appeared in both the retraction and comment datasets.

Because PubPeer comments vary in severity, we conducted a manual validation step to confirm that commented papers were a meaningful component of the problematic-paper set. After excluding the $2,617$ papers that had already been retracted, we randomly sampled $100$ commented papers for inspection. Of these, $95$ involved concerns related to image manipulation or duplication, data fabrication or falsification, ethical approval, or informed consent. Seven commented papers were retracted after the observation period. On this basis, we treated both retracted papers and commented papers as problematic papers throughout the study.

\textit{Paper-mill papers}. We further identified a subset of retracted papers associated with paper mills. These were identified from two sources: online lists of suspicious papers posted on blogs or social media platforms, and retraction notices that explicitly indicated third-party involvement or paper-mill-like behavior. Indicative phrases in retraction notices included ``authorship for sale", ``suspicious changes in authorship", ``manipulation of authorship", ``email addresses associated with multiple researcher accounts", ''carried out by a third party", ``third-party involvement", and ``similarities with (un)published articles from a separate third-party institute" \cite{JDIS:Amend}. Using these criteria, we identified $1,165$ of the $2,961$ retracted papers as paper-mill-related.

\textit{Reasons for retraction or comment}.
Retraction notices and PubPeer comments typically outline the reasons for retraction or the concerns raised, providing a natural basis for classification. Drawing on the classification framework established in our prior work \cite{JDIS:Amend}, Table \ref{tab:reasons} reports the reasons identified for $2,961$ retracted papers and $6,491$ commented papers. Because PubPeer comments have not been verified through formal investigations, many common retraction categories do not yet apply. Moreover, some comments merely question the reliability of article content without establishing whether the underlying issues are intentional or unintentional; such cases are therefore classified as ``unreliable".

\begin{table}[ht]
\caption{Frequency of identified reasons for retractions and comments.}
    \centering
    \begin{tabular}{ccc}
    \hline
         Reasons & Retracted papers & Commented papers\\ \hline
        Image Duplication/Fabrication\\/Manipulation/Plagiarism & $1,631$ & $4,634$ \\    
        Data Duplication/Fabrication\\/Manipulation/Plagiarism & $690$ & $576$ \\ 
        Inappropriate Authorship & $64$ & N/A \\ 
        Ethical Violations & $89$ & $72$ \\ 
        Paper Mills & $1,165$ & $697$ \\
        Fake Peer Review & $365$ & N/A \\ 
        Other Misconduct & $106$ & $50$ \\  \hline
        Error & $298$ & N/A \\
        Unreliable & N/A & $1,457$ \\  \hline
    \end{tabular}
    \label{tab:reasons}\\
    \textit{Please note that each article may be associated with multiple reasons for retraction or comment.}
\end{table}

\subsection{Identification of specific ncRNAs} \label{Identification}
\textit{microRNAs}. We identified individual microRNAs using miRetrieve, an R package and web application developed for microRNA text mining \cite{NAR:miRetrieve}. miRetrieve uses regular expressions to detect variant spellings of microRNA names in text and normalize them to a standardized format. We applied this approach to titles and abstracts in the ncRNA corpus. Using this procedure, approximately $1,700$ distinct microRNAs were identified across $71,896$ papers, most of which belonged to the MicroRNA micro-topic (Table \ref{tab:microtopics}).

\textit{lncRNAs}.  To identify individual lncRNAs, we first compiled names and name variants from several reference resources, including LNCipedia \cite{NAR:LNCipedia}, GENCODE \cite{NAR:GENCODE}, HGNC \cite{NAR:HGNC}, and LncRNADisease \cite{NAR:LncRNADisease}. We standardized these names and then matched them against article titles and abstracts to identify referenced lncRNAs. This procedure identified approximately $3,600$ distinct lncRNAs across $27,985$ papers, most of which belonged to the lncRNA micro-topic (Table \ref{tab:microtopics}).

Our entity-identification strategy was designed to capture standardized names and known variants, but it may not cover all possible spellings used in the literature. We therefore interpret these counts as conservative estimates. We also note that, unlike microRNAs and lncRNAs, exosome-related terminology and RNAi nomenclature do not have comparably unified reference resources. For this reason, exosome- and RNAi-related entities were not identified by name in the present study.

\begin{table}[ht]
\caption{Distribution of Individual (Indiv.) lncRNAs and Individual microRNAs groups identified by ncRNA names across micro-topics of the ncRNA field, along with their respective numbers of retracted and commented papers.}
    \centering
    \begin{tabular}{cccccccc}
    \hline
    \multirow{2}{3em}{Category} & \multirow{2}{3em}{Total} & \multicolumn{4}{c}{Micro-Topics} &\multirow{2}{3em}{Retracted} &\multirow{2}{3em}{Commented}  \\
    
    & & MicroRNA & lncRNA & Exosomes & RNAi &  & \\ 
    \hline  
    ncRNA &$153,943$ & $72,276$ &$46,131$ &	$25,748$ &	$9,788$ &	$2,961^\#$ & $9,108$  \\
    \textit{Indiv. microRNAs*}& $71,896$ & $51,576$ & $16,640$ &	$3,553$ & $127$ & $2,399$ & $6,943$		 \\
    \textit{Indiv. lncRNAs*}& $27,985$ &	$3,980$ &$22,910$ &	$886$ & $209$ &$1,129$ & $3,303$\\			
    
    \hline    
    \end{tabular}     
    \label{tab:microtopics}
    
    * \textit{Because a single paper can be classified into both the Individual microRNAs and Individual lncRNAs groups based on ncRNA names, there are $14,231$ overlapping papers between these two categories.} \# \textit{Among the retracted papers, $1,165$ are attributed to paper mills.}
 \end{table}

\subsection{Textual similarity}
To assess large-scale textual similarity, we used the published two-dimensional map of the biomedical literature developed from PubMed abstracts using PubMedBERT embeddings and t-SNE dimensionality reduction \cite{Patterns:landscape}. In this representation, shorter distances between papers indicate greater similarity in abstract-level textual content. Using PMIDs, we linked ncRNA papers to their coordinates on this map and then examined the distribution of problematic papers within the projected similarity space.

\subsection{Assessment of potential image duplication}
Potential image duplication was assessed using FigCheck, an automated image-screening system available at \url{https://www.figcheck.com/}. FigCheck applies neural-network-based detection and automated annotation to identify regions of suspected duplication within figures. In this study, FigCheck outputs were used as indicators of potential duplication rather than as definitive proof of misconduct.

\subsection{Organization type classification}
Institutional organization types were obtained from Dimensions, which categorizes affiliations as Education, Healthcare, Facility, Nonprofit, Government, Company, or other types. Because many university-affiliated hospitals are coded as Education in Dimensions, we manually reassigned institutions with hospital affiliations to the Healthcare category for the purposes of this study.

We then classified institutions into two groups: Healthcare and Non-Healthcare. A paper was classified as a Healthcare paper if at least one affiliated institution belonged to the Healthcare group; otherwise, it was classified as a Non-Healthcare paper. Using this rule, $91,455$ papers were classified as Healthcare papers and $62,488$ as Non-Healthcare papers. When calculating institutional publication volume, Healthcare institutions were credited only for Healthcare papers, and Non-Healthcare institutions were credited only
for Non-Healthcare papers.

\section{Results}

\subsection{Concentration of problematic papers in under-investigated ncRNAs}

\begin{figure}[ht]
    \centering
    \includegraphics[width=1\linewidth]{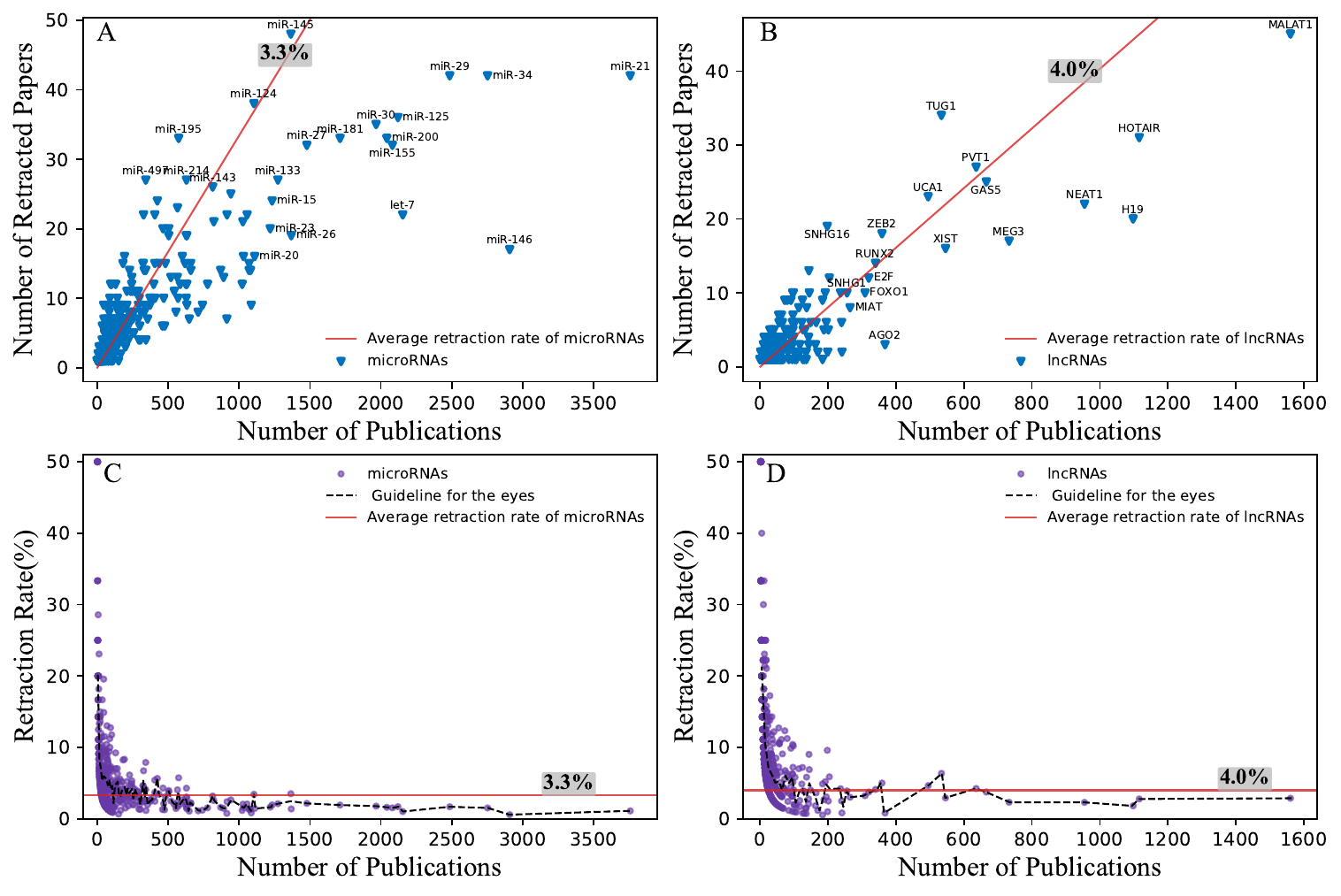}
    \caption{Analysis of retracted papers related to specific microRNAs and long non-coding RNAs (lncRNAs). Panel (A) shows the count of retracted papers in relation to the number of publications for specific microRNAs, while panel (B) presents the same for lncRNAs. Panels (C) and (D) illustrate the corresponding retraction rates relative to the number of publications for specific microRNAs and lncRNAs, respectively. The solid lines represent the average retraction rates for the microRNA and lncRNA categories. The dashed line serves as a visual guideline.} 
    \label{fig:RNA_retracted}
\end{figure}

We first examined whether problematic papers were concentrated in specific named ncRNAs. Among the $71,896$ papers that referenced individual microRNAs, $2,399$ were retracted, including $994$ linked to paper mills, corresponding to a retraction rate of $3.3\%$ (Fig. \ref{fig:RNA_retracted}A). Among the $27,985$ papers that referenced individual lncRNAs, $1,129$ were retracted, including $430$ linked to paper mills, corresponding to a retraction rate of $4.0\%$ (Fig. \ref{fig:RNA_retracted}B). Both rates were above the field-wide ncRNA retraction rate of $1.92\%$ \cite{bioRxiv:ncRNA}.

Retraction rates declined as the number of publications on a named ncRNA increased (Fig. \ref{fig:RNA_retracted} C and D). Among microRNAs and lncRNAs with at most $10$ publications, retraction rates reached $20.2\%$ ($47/233$) and $21.3\% $ ($132/619$), respectively. By contrast, more frequently studied ncRNAs showed lower rates. For example, miR-195 and TUG1 had $574$ and $534$ publications, respectively, with retraction rates of $5.7\%$ and $6.4\%$. Highly prevalent ncRNAs such as miR-146, miR-34, and NEAT1 remained below the field average.

\begin{figure}[ht]
    \centering
    \includegraphics[width=1\linewidth]{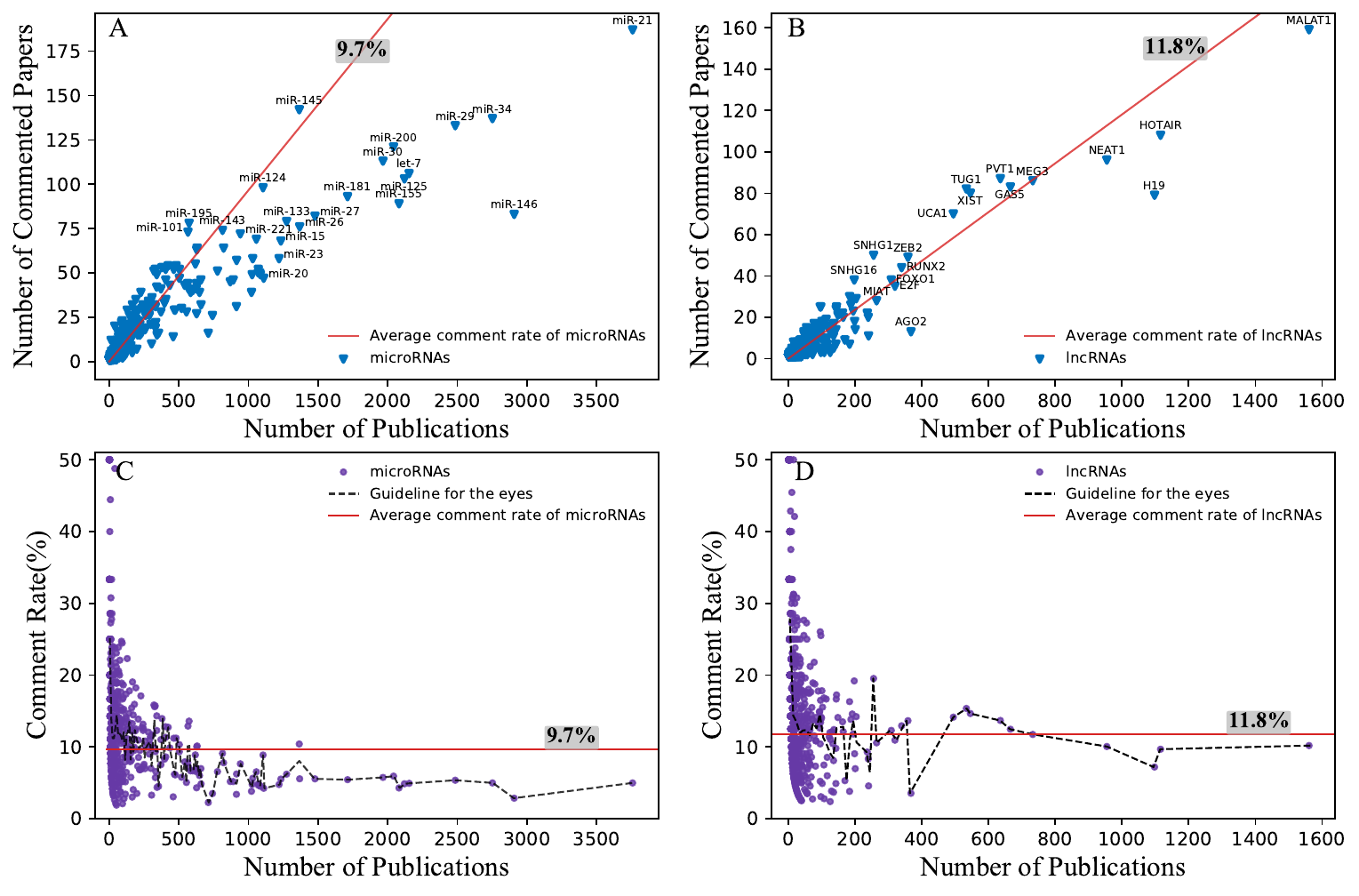}
    \caption{Analysis of commented papers related to specific microRNAs and long non-coding RNAs (lncRNAs). Panel (A) shows the count of commented papers in relation to the number of publications for specific microRNAs, while panel (B) presents the same for lncRNAs. Panels (C) and (D) illustrate the corresponding comment rates relative to the number of publications for specific microRNAs and lncRNAs, respectively. The solid lines represent the average comment rate.} 
    \label{fig:RNA_commented}
\end{figure}

A similar pattern was observed for commented papers. Of the papers referencing individual microRNAs, $6,943$ received PubPeer comments, yielding a comment rate of $9.7\%$. Of the papers referencing individual lncRNAs, $3,303$ received comments, yielding a comment rate of $11.8\%$. Both rates exceeded the field average of $5.92\%$ \cite{bioRxiv:ncRNA}. Among microRNAs and lncRNAs with fewer than $10$ publications, comment rates rose to $25.6\%$ ($150/587$) and $27.9\%$ ($423/1,515$), respectively (Fig. \ref{fig:RNA_commented}). Overall, both retraction and comment rates decreased as publication volume increased.

\subsection{Clustering of problematic papers in high-similarity regions}

\begin{figure}[ht]
    \centering
    \includegraphics[width=0.95\linewidth]{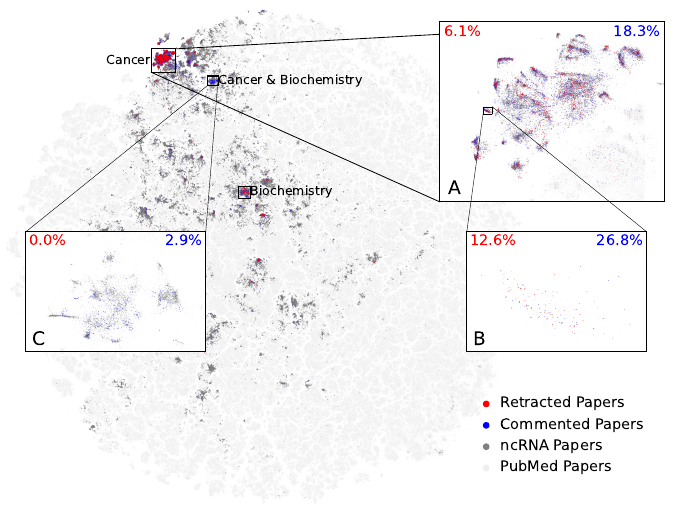}
    \caption{ 2D map for ncRNA papers. All ncRNA papers are displayed against the backdrop of PubMed papers, with approximately $2,900$ retracted papers highlighted in red and over $9,000$ commented papers highlighted in blue. The labels for the regions are based on the frequency of their occurrence. Inset A highlights regions labeled \textit{Cancer}, which exhibit a higher density of retracted papers ($6.1\%$) and commented papers ($18.3\%$), particularly in relation to microRNA research. Subinset B identifies a specific area with an even greater proportion of retracted papers and commented papers with title formats A and B in Eq. (\ref{Title}). Inset C  highlights the \textit{Cancer \& Biochemistry} region , in which many commented papers share title format C in Eq. (\ref{Title}). }
    \label{fig:sciencemap}
\end{figure}

We next projected ncRNA papers onto the published two-dimensional biomedical similarity map \cite{Patterns:landscape}. Problematic papers were not uniformly distributed across this space. Instead, they were concentrated in specific regions, particularly within the area labeled \textit{Cancer}. This region contained $30,339$ ncRNA papers, including $1,863$ retracted papers, for a retraction rate of $6.1\%$, and $5,571$ commented papers, for a comment rate of $18.3\%$ (Fig. \ref{fig:sciencemap}A). Among the retracted papers in this region, $603$ were linked to paper mills.

Within this broader cancer region, problematic papers were especially concentrated in subsets characterized by recurrent title structures. In the area highlighted in Fig. \ref{fig:sciencemap}B, we identified $340$ papers following structured title formats among a total of $452$ publications. The dominant formats are summarized below:
\begin{equation} 
\begin{array}{ll}
\text{Format A}: & \text{MicroRNA-X does Y by (through, via) doing Z.} \\
\text{Format B}: & \text{MicroRNA-X does (be) Y.} \\
\text{Format C}: & \text{(Role, Mechanism, Impact, Function of) X RNAs in Y.}
\end{array}\label{Title}
\end{equation}
Across the broader cancer domain in Fig. \ref{fig:sciencemap}A, $7,525$ papers followed title format A, of which $546$ ($7.3\%$) were retracted and $1,624$ ($21.6\%$) were commented. A further $1,791$ papers followed title format B, of which $86$ ($4.8\%$) were retracted, and $291$ ($16.2\%$) were commented. Similar patterns were observed when microRNA was replaced by lncRNA or circRNA. For example, among $5,070$ lncRNA papers with these structured titles, $382$ ($7.5\%$) were retracted and $1,097$ ($21.6\%$) were commented (Table \ref{tab:titles}).

We also examined a set of $325$ non-retracted and non-commented papers from the high-similarity region in Fig. \ref{fig:sciencemap}B. FigCheck flagged $124$ of these papers ($38.1\%$) as containing suspected image duplication within individual papers. In addition, many papers in these clusters showed irregular author email patterns. Furthermore, $416$ of $452$ articles ($92.0\%$) include \textit{osteosarcoma} in their titles, a striking overrepresentation relative to the \textit{Cancer} area overall ($4.7\%$). These observations indicate that suspicious signals extend beyond the set of already retracted or commented papers.

\begin{table}[ht]
    \centering
    \caption{Retractions and comments associated with recurrent title formats in regions high-
lighted on the textual-similarity map.}
    \begin{tabular}{ccccc}
        \hline
         Region/Format & Papers & Retracted& Commented& Reference in Eq. \ref{Title} \\ \hline
         Fig. \ref{fig:sciencemap}A/Title A & $7,525$ & $546$ ($7.3\%$)& $1,624$ ($21.6\%$) & Format A  \\          
         Fig. \ref{fig:sciencemap}A/Title B & $1,791$ & $86$ ($4.8\%$)& $291$ ($16.2\%$) & Format B \\ 
         
         Fig. \ref{fig:sciencemap}A/Title A/B* & $5,070$ & $382$ ($7.5\%$) & $1,097$ ($21.6\%$) & Format A/B* \\
         
         Fig. \ref{fig:sciencemap}B/Title A & $296$ & $40$ ($13.5\%$) & $87$ ($29.4\%$) & Format A \\ 
         Fig. \ref{fig:sciencemap}B/Title B & $34$  & $6$ ($17.6\%$) & $13$ ($38.2\%$) & Format B \\ 
         
         Fig. \ref{fig:sciencemap}C/Title C & $6,330$ & $3$ ($0.05\%$)& $206$ ($3.2\%$) & Format C \\ \hline
    \end{tabular} 
    \label{tab:titles}
    
    \textit{*The titles for lncRNA analogs follow an A/B format  in Eq.\ref{Title}.}
\end{table}

The map also distinguished regions with different profiles of concern. The \textit{Cancer \& Biochemistry} region contained $8,522$ papers, only four retractions, and $248$ commented papers, yielding a comment rate of $2.9\%$ (Fig. \ref{fig:sciencemap}C). More than $200$ of these $248$ commented papers followed the descriptive title pattern labeled Format C in Eq. \ref{Title}, and $6,330$ papers in this region followed the same format (Table \ref{tab:titles}). By contrast, the Biochemistry region contained $6,234$ papers, $181$ retractions ($2.9\%$), including $92$ linked to paper mills, and $499$ commented papers ($8.0\%$).

\subsection{Overrepresentation of problematic papers in healthcare institutions}

\begin{figure}[ht]
    \centering
    \includegraphics[width=1.0\linewidth]{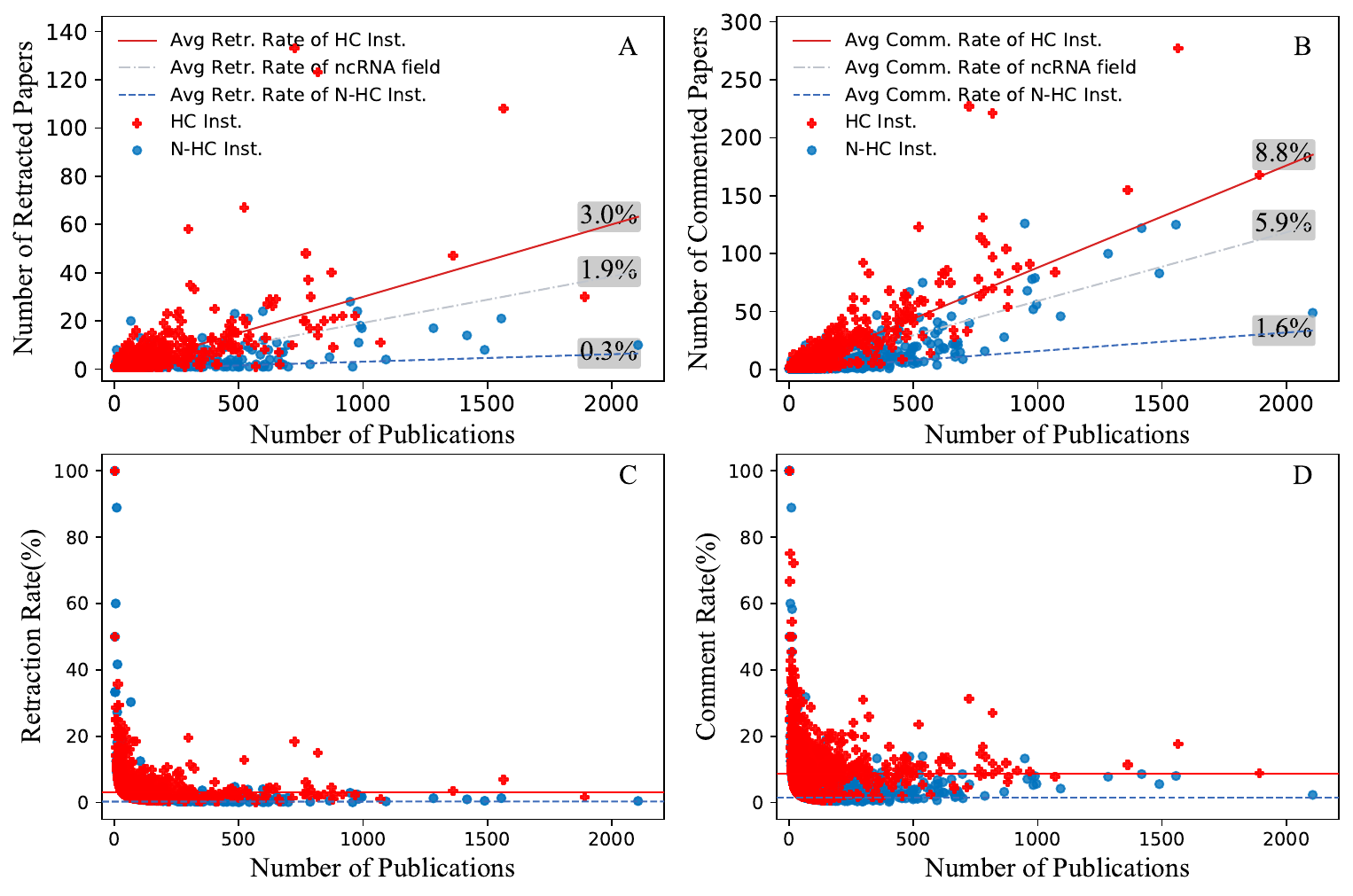}
    \caption{ Trends in retracted and commented papers related to healthcare (HC) and non-healthcare (N-HC) institutions (Inst). Panel (A) displays the total number of publications and retracted papers, while panel (B) presents the commented papers on PubPeer. Panel (C) illustrates the corresponding retraction rate, while panel (D) illustrates the corresponding comment rates relative to the number of publications. The solid, dashed, and dotted-dashed lines represent the average (Avg) retraction or comment rates for papers from healthcare institutions, non-healthcare institutions, and the ncRNA field, respectively. }
    \label{fig:institute}
\end{figure}

Institutional patterns were similarly uneven. Of the $153,943$ ncRNA papers in the corpus, $91,455$ ($59.4\%$) were affiliated with healthcare institutions. These institutions accounted for $2,769$ of the $2,961$ retracted papers ($93.5\%$), corresponding to a retraction rate of approximately $3.0\%$. By contrast, the $62,488$ papers from non-healthcare institutions accounted for only $192$ retractions, corresponding to a retraction rate of $0.3\%$ (Fig. \ref{fig:institute}A).

Healthcare institutions also dominated the set of commented papers. They accounted for $8,093$ commented papers ($88.9\%$), yielding a comment rate of $8.8\%$. Non-healthcare institutions accounted for $1,015$ commented papers, corresponding to a comment rate of approximately $1.6\%$ (Fig. \ref{fig:institute}B).

Within the healthcare sector, substantial heterogeneity was observed across institutions. Many low-output institutions showed very high retraction and comment rates, in some cases exceeding $20\%$, whereas higher-output institutions generally showed lower rates (Fig. \ref{fig:institute}C and D). Geographically, retracted papers originated from $31$ countries or regions, and commented papers from $66$. China differed sharply from other countries: healthcare institutions accounted for $97.5\%$ of retracted papers and $95.4\%$ of commented papers in China, compared with $44.6\%$ and $60.9\%$ in the United States. When China was excluded, healthcare institutions accounted for only $20.4\%$ of retracted papers and $40.1\%$ of commented papers in the remaining countries (Table \ref{tab:Country}). Similarly, healthcare institutions accounted for the majority of papers in categories related to name identification.

\begin{table}[ht]
\caption{Stratification of retracted and commented papers by country of origin, ncRNA names, and institution type.}
    \centering
    \begin{tabular}{cccccccccc}
    \hline
    \multicolumn{2}{c}{\multirow{2}{*}{Category}} & \multirow{2}{3em}{Papers} & \multirow{2}{3em}{Health} & \multicolumn{3}{c}{Retracted papers} &\multicolumn{3}{c} {Commented papers}  \\
    
    & &  &  &  Papers & Health & Percent & Papers & Health & Percent \\
    \hline
     Field & ncRNA & $153,943$ & $91,455$ & $2,961$ & $2,769$ & $93.5\%$ & $9,108$ & $8,093$ & $88.9\%$\\    
    \hline
    & China & $84,748$ & $70,104$ & $2,809$ & $2,738$ & $97.5\%$ & $8,029$ & $7,660$ & $95.4\%$ \\
    %\hline
    & USA	& $29,462$ & $11,094$ & $101$ & $45$ & $44.6\%$ & $787$ & $479$ & $60.9\%$ \\
    %\hline
    \multirow{1}{2em}{Country} & Iran & $4,024$ & $749$ & $23$ & $7$ & $30.4\%$ & $180$ & $67$ & $37.2\%$ \\
    %\hline
    & Japan & $5,109$ & $1,417$ & $14$ & $4$ & $28.6\%$ & $56$ & $26$ & $46.4\%$ \\
    %\hline
    & Italy & $5,674$ & $2,571$ & $12$ & $4$ & $33.3\%$ & $146$ &	$76$ & $52.1\%$ \\
    \hline

   \multirow{2}{2em}{Name} & \textit{Indiv. microRNAs} & $71,896$ & $52,938$ & $2,399$ & $2,275$ & $94.8\%$ & $6,943$ & $6,388$ & $92.0\%$ \\
    %\hline
    & \textit{Indiv. lncRNAs} & $27,985$ & $23,159$ & $1,129$ & $1,090$ & $96.5\%$ & $3,303$ & $3,130$ & $94.8\%$ \\
    \hline
    \end{tabular}     
    \label{tab:Country}
 \end{table}

\subsection{Disproportionate share from a small set of journals}
The ncRNA corpus included $153,943$ papers published in approximately $5,000$ journals from nearly $800$ publishers. By contrast, the retracted and commented papers were concentrated in a much smaller subset of outlets. Retracted papers came from more than $300$ journals across over $50$ publishers, whereas commented papers appeared in about $700$ journals from more than $100$ publishers.

\begin{figure}[ht]
    \centering
    \includegraphics[width=0.99\linewidth]{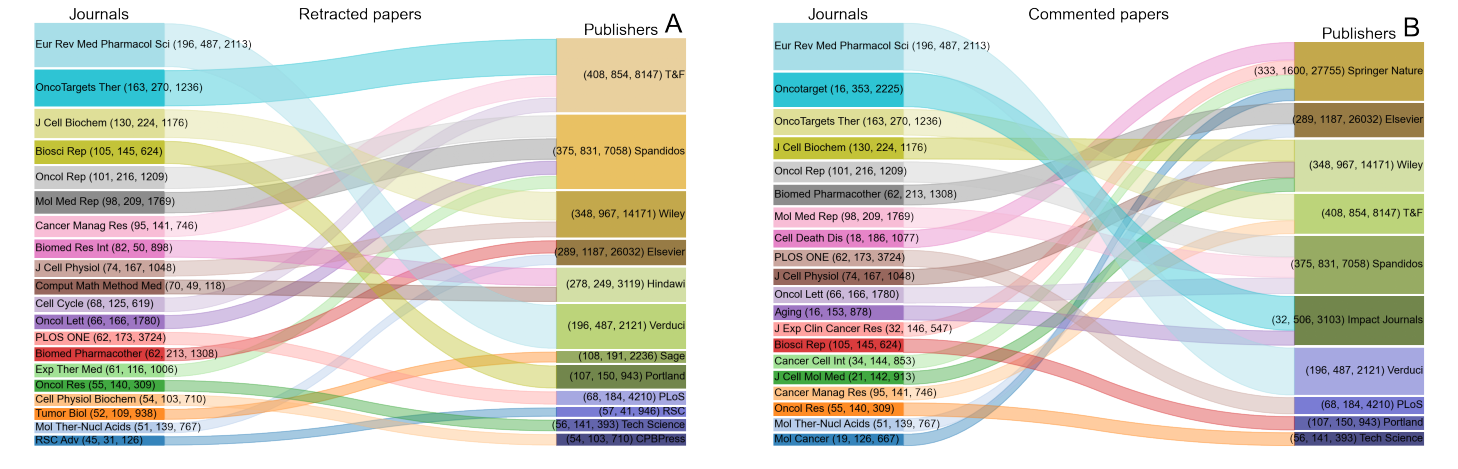} %\includegraphics[width=0.49\linewidth]{Figure_5B.eps}
    \caption{Retracted and commented papers in leading journals and publishers. Panel (A) shows the number of retracted articles, while panel (B) presents the number of commented articles for the top $20$ journals and their respective publishers. The numbers in parentheses indicate the counts of retracted and commented articles, along with the total number of publications in the ncRNA field. }
    \label{fig:publisher}
\end{figure}

The top $20$ journals by retraction count published $22,224$ papers ($14.4\%$ of the corpus) but accounted for $1,690$ retractions ($57.1\%$ of all retracted papers), corresponding to a retraction rate of $7.6\%$. The $12$ publishers associated with these journals accounted for $2,344$ retractions ($79.2\%$ of all retracted papers) while publishing $70,086$ ncRNA papers ($45.5\%$ of the corpus), corresponding to a retraction rate of $3.3\%$ (Fig. \ref{fig:publisher}A). For commented papers, the top $20$ journals published $24,969$ papers ($16.2\%$ of the corpus) and accounted for $3,940$ commented papers ($43.3\%$ of all commented papers), corresponding to a comment rate of $15.8\%$. The $10$ publishers associated with these journals published $93,933$ ncRNA papers ($61.0\%$ of the corpus) and accounted for $6,907$ commented papers ($75.8\%$ of all commented papers), corresponding to a comment rate of $7.4\%$ (Fig. \ref{fig:publisher}B).

\begin{figure}[ht]
    \centering
    \includegraphics[width=1\linewidth]{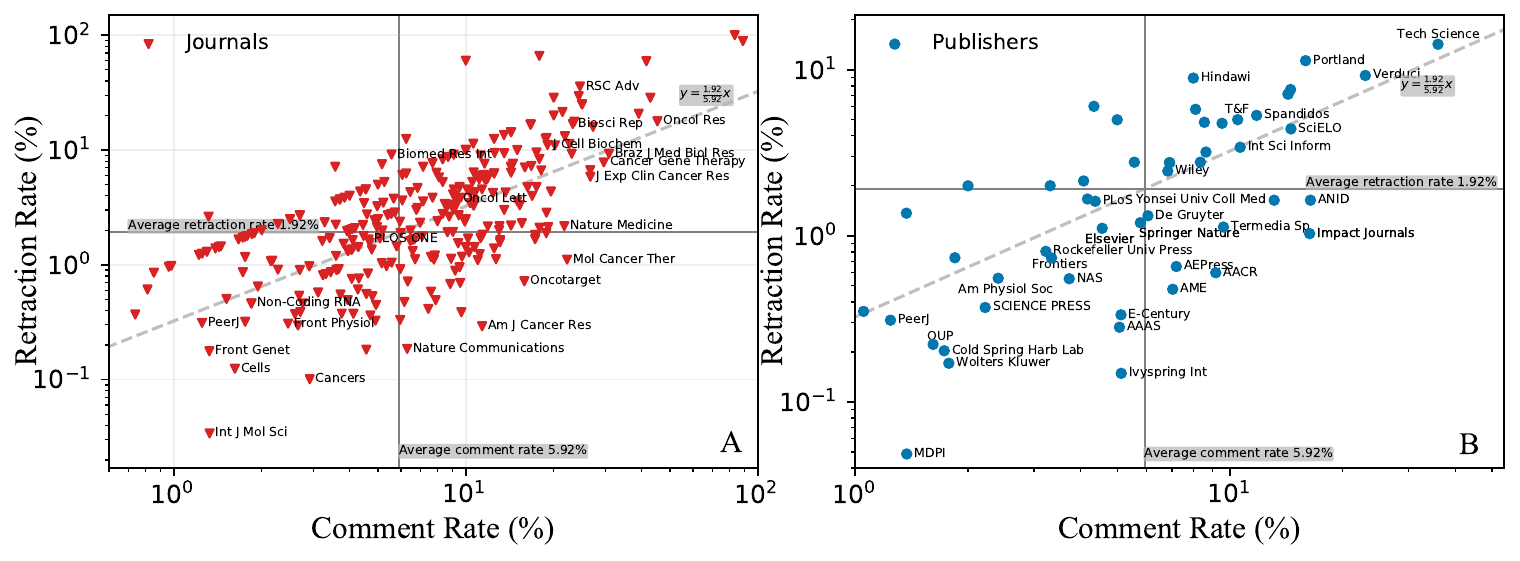}
    \caption{Retraction rates in relation to comment rates for journals and publishers. Panel (A) displays data for journals, while panel (B) focuses on publishers. The solid horizontal and vertical lines represent the retraction rate and comment rate for the ncRNA field, respectively. The dashed line illustrates that the average Retraction-Comment Ratio of the ncRNA field, specifically described by the equation $y=\frac{1.92}{5.92}x$.}
    \label{fig:retractionrate}
\end{figure}

Journal and publisher responsiveness to commented papers also varied substantially. We defined the Retraction-Comment Ratio (RCR) as the ratio of retraction rate to comment rate. For the ncRNA field as a whole, the average RCR was $1.92/5.92\approx0.324$. Many journals and publishers fell below this benchmark in Fig. \ref{fig:retractionrate}, indicating lower retraction activity relative to the number of commented papers. For example, \textit{Oncotarget} published $2,225$ ncRNA papers but had only $16$ retractions and $353$ commented papers, yielding an RCR of $0.045$. MDPI journals in this field had only four retracted papers and $113$ commented papers, yielding an RCR of $0.035$.

\section{Discussion and Conclusion}
\subsection{Discussion}
Post-publication peer review platforms such as PubPeer have emerged as vital alert systems, enabling publishers and journals to rapidly identify articles of questionable reliability, investigate them, and ultimately retract those substantiated as unreliable \cite{JASIST:PubPeer,RIPR:peerreview}. Retraction functions as a critical self-correction mechanism in science, cautioning researchers against citing unreliable literature \cite{PLOSONE:Self-correction,RIPR:Spread}. However, the current system fails to effectively mitigate the persistence of problematic papers. Only a small fraction of commented papers have been retracted, while retracted papers continue to be persistently cited \cite{PBR:lives,PLOSONE:Self-correction}. This suggests that these problematic papers continue to erode the foundation for future innovation. Consequently, developing a comprehensive understanding of the characteristics of problematic papers and establishing methods for their timely identification have become critical priorities.

This study presents a systematic analysis of the characteristics of retracted and commented papers within the ncRNA field. The results reveal that problematic papers are not randomly distributed. Notably, they are concentrated along four dimensions: under-investigated ncRNAs, regions of high textual similarity, healthcare institutions, and a relatively small set of journals. These patterns indicate that the integrity problems observed in ncRNA are structured rather than isolated, and that a substantial number of potential issues may remain undetected within the literature. 

First, the marked inverse association between publication volume and both retraction and comment rates suggests that under-investigated ncRNAs are particularly vulnerable to problematic publication. When few studies exist on a named ncRNA, claims may be harder to evaluate because there is less accumulated domain knowledge against which to assess plausibility, methods, and results. This knowledge gap may further impede peer reviewers' ability to adequately evaluate study quality, thereby increasing the risk of problematic papers going undetected. Thus, research on under-investigated ncRNAs is particularly susceptible to low-quality and fraudulent work \cite{BIOMARKER:fraudulent}. This is consistent with previous work showing that poorly characterized genes and nucleotide-sequence reagents can become recurrent targets of unreliable research \cite{PLOSBIOLOGY:under-investigate, NRMC:lncRNA,LSA:wrongly}. Furthermore, a significant amount of ncRNA research has employed wrongly identified nucleotide sequence reagents as targeting agents. For example, numerous studies on miR-145 have incorporated one or more inaccurately identified nucleotide sequences, leading to unreliable conclusions \cite{LSA:wrongly}.

Second, textual-similarity analyses show that problematic papers tend to cluster in recognizable regions of the literature, sharing recurrent title structures and significant textual similarities in their abstracts. Beyond these textual patterns, many non-retracted and non-commented papers display suspected image duplication or irregular email addresses. Additionally, retracted and commented papers show close topological proximity within the citation network. The coexistence of repeated title patterns, high retraction and comment rates, suspected image duplication, irregular email addresses, and paper-mill-linked retractions suggests that some subsets of the ncRNA literature may have been shaped by standardized or template-based production characteristic of paper mill operations \cite{JCE:papermill,BIOMARKER:fraudulent}. However, similarity alone should not be treated as proof of misconduct. Rather, it serves as a screening signal that can help identify clusters of papers warranting closer examination \cite{Patterns:landscape}.

Third, the institutional analysis reveals that healthcare institutions are responsible for a disproportionate share of problematic papers in the ncRNA field, a pattern most pronounced in China, where healthcare institutions account for a substantial portion of the problematic-paper set. Notably, institutions with low output exhibit higher retraction and comment rates compared to their larger counterparts. Although the data do not establish causation, this pattern is consistent with prior literature linking publication pressure, incentive structures, and limited research training in hospital settings to vulnerability to paper-mill involvement \cite{NATURE:papermill,NATURE:hospital,NATURE:CASHREWARDS}. In Chinese hospitals, physicians are required to publish a specific number of papers to qualify for promotion \cite{NATURE:papermill,NATURE:hospital} and have historically received financial bonuses for publications \cite{NATURE:CASHREWARDS}, yet many, especially those in primary care, lack sufficient training in rigorous methodologies, research integrity, and ethics \cite{NATURE:hospital}. These systemic pressures and knowledge gaps contribute to questionable research practices and render some physicians vulnerable to exploitation by paper mills \cite{NATURE:papermill,NATURE:hospital}, underscoring research integrity as a pressing challenge for Chinese hospitals.

Fourth, problematic papers are concentrated in a limited number of journals and publishers, and many outlets show low Retraction-Comment Ratios relative to the field average. This suggests that journals differ not only in their exposure to problematic submissions but also in how actively they respond once concerns are raised. This disparity likely reflects differences in editorial standards: some journals prioritize publication quantity over quality control, rendering them vulnerable targets for paper mills \cite{PNAS:editor}. Indeed, individuals affiliated with paper mills may impersonate well-known scholars to obtain positions on editorial boards, thereby compromising the integrity of peer review and gaining the power to manipulate editorial decisions \cite{Science:editorial}. Consequently, paper mills not only produce large volumes of low-quality, fraudulent papers but also facilitate their publication in specific journals, leading to a substantial increase in annual publication volume accompanied by a surge in retractions and PubPeer comments. When such journals lose favor with paper mills due to reputational damage, their annual publication numbers decline sharply, as evidenced by J. Cell. Biochem. and Biomed Res. Int. The persistence of large numbers of commented but non-retracted papers further indicates that post-publication correction remains incomplete.

Overall, these findings suggest that a substantial number of problematic papers may remain undetected in the ncRNA literature. More importantly, they show that the combination of topic-level concentration, similarity-based clustering, institutional concentration, and journal-level accumulation can help prioritize surveillance efforts toward high-risk segments of the literature.

\subsection{Implications for research integrity}
The findings carry several implications for research integrity practice. While identifying and containing problematic papers presents formidable challenges, addressing them is imperative. This necessitates a comprehensive, multi-stakeholder framework in which the primary objective is to identify and retract problematic papers rather than to assign blame \cite{PNAS:editor}.

A coordinated effort requires all stakeholders to fulfill their distinct responsibilities. Researchers should promptly disclose identified fraudulent papers, e.g., by posting comments on PubPeer, and refrain from citing questionable works. Research-integrity experts should analyze the characteristics of problematic papers, develop detection methods \cite{PNAS:editor}, and compile lists of suspicious publications for surveillance. Institutions, particularly those in clinical settings, should strengthen training in research methods, data management, and research integrity standards. Funding agencies should reduce incentives that reward publication volume without sufficient regard for quality, while implementing policies that encourage the disclosure of previously flawed work and protect against severe penalties.

For journals and publishers, the concentration of problematic papers in a limited set of outlets and the low Retraction-Comment Ratios observed for some publishers highlight the need for stronger editorial screening, faster post-publication investigation, and more transparent correction processes. Before publication, journals should implement strict controls on publication volume, conduct rigorous vetting of editorial board members and guest editors to prevent infiltration by paper-mill operatives \cite{PNAS:editor}, leverage expert reviewers to detect fraudulent submissions \cite{JOI:peerrivew}, and integrate automated forensics tools for image duplication and template-based text detection. Specific measures include multi-layered identity verification protocols, such as mandatory ORCID linkage and institutional email verification, as well as mandatory raw data submission, image-anomaly detection, and template-based text screening.

After publication, systematic post-publication audits  should complement these pre-publication measures by employing AI-driven similarity checks to identify and promptly retract clustered problematic papers. Furthermore, journals bear the responsibility to actively resolve any papers flagged for concern, carry out timely investigations, and ensure that the findings are made publicly available. At the publisher level, collaborative development of shared AI detection tools across multiple journals would enhance cost-effectiveness and coverage. Equally important, sharing submission metadata across journals/publishers would enable cross-checking for overlapping manuscripts, thereby facilitating the identification of paper-mill operations that submit the same work to multiple outlets simultaneously.

However, even well-coordinated retraction efforts address only part of the challenge. A more persistent difficulty lies in purging erroneous knowledge once it has been integrated into the scientific record. Field experts should collaborate to annotate and flag false claims within knowledge graphs, thereby preventing subsequent research from building on inaccurate foundations. The broader community, in turn, should systematically re-evaluate published works that cite or depend on these identified inaccuracies.

\subsection{Limitations}

First, our analysis focuses on the ncRNA field as defined through citation-linkage clustering. Although this enables in-depth examination of a specific domain, the extent to which our findings generalize to other disciplines remains uncertain. In addition, because topic identification relied on citation linkages, some relevant ncRNA publications may have been omitted. Future work should extend this approach to additional topics and employ large language models or other semantic methods to improve topic delineation.

Second, our identification of microRNAs and lncRNAs was based on matching standard names and name variants in titles and abstracts. This surface-level strategy may have missed relevant papers. Full-text analysis would allow more comprehensive identification of specific ncRNAs.

Third, the dataset was drawn exclusively from Web of Science. Publications indexed only in other major databases, such as Scopus, PubMed, or OpenAlex, were therefore excluded, and our estimates may understate the full extent of problematic papers.

Finally, our textual-similarity analysis was limited to titles and abstracts. This constraint may miss more complex forms of manipulation. Future studies should incorporate full-text analysis to improve detection.

\subsection{Conclusions}
In conclusion, this study documents the substantial infiltration of problematic publications into the ncRNA field. By examining specific ncRNA entities, textual similarity, affiliation patterns, and journal concentration, we provide a multifaceted characterization of these papers and show how integrity risks can accumulate within a rapidly expanding domain. Although our analysis is confined to ncRNA, the need for transparency, rigorous verification, and timely correction extends across disciplines. The specific manifestations of the problem may vary by field, but the analytical framework developed here can inform broader efforts to detect and counter research fraud. Safeguarding the scientific record requires coordinated, cross-disciplinary action by researchers, journals, institutions, and publishers.

\section*{Acknowledgments}
This study is partially supported by the LIS Outstanding Talents Introducing Program, Bureau of Development and Planning, CAS (2022), the Beijing Natural Science Foundation (grant no. 9242006), and the National Natural Science Foundation of China (grant no. 71974017).

\section*{CRediT authorship contribution statement}
Ying Lou:Data curation, Visualization, Formal analysis, Writing – original draft. Zhengyi Zhou: Data curation, Formal analysis, Visualization. Guosheng Wang: Data curation, Formal analysis, Methodology. Zhesi Shen: Conceptualization, Methodology, Visualization, Writing – review \& editing. Menghui Li: Conceptualization, Formal analysis, Visualization, Writing – review \& editing.

\section*{Disclosure statement}
There are no competing interests to declare. 

\section*{Declaration of generative AI and AI-assisted technologies in the writing process}
During the preparation of this work, the authors used ChatGPT in order to improve readability and language. After using this tool, the authors reviewed and edited the content as needed and take full responsibility for the content of the publication.

% \bibliographystyle{cas-model2-names}
% \bibliography{ncrna-reference}

\end{document}